# Choosing a Suitable Requirement Prioritization Method: A Survey

Esraa Alhenawi[1], Shatha Awawdeh[2], Ruba Abu Khurma[3],
Maribel García-Arenas[4], Pedro A. Castillo[4], Amjad Hudaib[5]

[1]*Faculty of Information Technology, Zarqa University, Zarqa, Jordan*
ealhenawi@zu.edu.jo

[2]*Applied Science Private University, Amman, Jordan, a_awawdeh@asu.edu.jo*

[3]*MEU Research Unit, Faculty of Information Technology, Middle East University, Amman, 11831, Jordan, rubaabukhurma82@gmail.com*

[4]*Department of Computer Engineering, Automation and Robotics, University of Granada, Granada, Spain, mgarenas@ugr.es, pacv@ugr.es*

[5]*King Abdullah II School for Information Technology, The University of Jordan, Amman, Jordan ahudaib@ju.edu.jo*

**Abstract**

*Software requirements prioritization plays a crucial role in software development. It can be viewed as the process of ordering requirements by determining which requirements must be done first and which can be done later. Powerful requirements prioritization techniques are of paramount importance to finish the implementation on time and within budget. Many factors affect requirement prioritization such as stakeholder expectations, complexity, dependency, scalability, risk, and cost. Therefore, finding the proper order of requirements is a challenging process. Hence, different types of requirements prioritization techniques have been developed to support this task. In this survey, we propose a novel classification that can classify the prioritization techniques under two major classes: relative and exact prioritization techniques class, where each class is divided into two subclasses. We depend in our classification on the way the value of ranking is given to the requirement, either explicitly as a specific value in the case of the exact prioritization techniques class, or implicitly in the case of the Relative prioritization technique class. An overview of fifteen different requirements prioritization techniques are presented and organized according to the proposed classification criteria's. Moreover, we make a comparison between methods that are related to the same subclass to analyze their strengths and weaknesses. Based on the comparison results, the properties for each proposed subclass of techniques are identified. Depending on these properties, we present some recommendations to help project managers in the process of selecting the most suitable technique to prioritize requirements based on their project characteristics (number of requirements, time, cost, and accuracy).*

**Keywords:** *Software requirement engineering, requirement prioritization techniques, relative prioritization techniques, exact prioritization techniques*

## 1. Introduction

Software engineering is concerned with developing high-quality software using software process models and reliable techniques [53], where high-quality means: usable, learnable, secure, available, and reliable software with high performance. Software needs a systematic way of development due to its complexity. Software management is the process that specifies the area of software development through the software process model, which is called the software development life cycle (SDLC). [1]

Requirements prioritization is one of many critical activities of requirements engineering contributing towards making good decisions for software systems. This process aims to determine the most important requirements and eliminate unnecessary requirements to finish implementation on time and within budget. Moreover, it plays a key role in case of planning system releases to decide which requirements to implement in each release, according to budget and time on one hand and customer expectations on the other hand. Requirements prioritization is a challenging task as it can be affected by many factors such as stakeholder expectations, risk [2, 3, 4] cost, time [4], complexity, dependency, scalability, technical value and requirement change [5, 6].

The selection of the right set of requirements has a key rule to satisfy all the needs of the stakeholders and maximize the business value of the product. Moreover, it is considered highly significant in the decision-making process. On the other hand, the wrong set of requirements has an adverse effect on the quality of the product and the cost of the modification of the system later as well. The value of the requirements is calculated by using a suitable requirement prioritization technique. Requirements prioritization techniques vary greatly. The selection of technique depends on the status of the project where each technique has its own advantages and disadvantages. Some techniques work well with projects having a large number of requirements, while others are unable to give acceptable results when dealing with a large number of requirements.

The research questions are as follows:
- What are the most popular requirement prioritization techniques?
- How can we classify requirement prioritization techniques based on the way the value of ranking is given to the requirement?
- How project managers can select the most suitable prioritization technique from the huge number of techniques, based on their project characteristics including three major directions including number of requirements, time, cost, and accuracy the project must satisfy?

The main contributions of this survey are summarized as follows:
- The survey presents some of the most popular requirement prioritization techniques (about fifteen techniques).
- A new classification of these techniques is proposed based on the way the value of ranking is given to the requirement, either explicitly as a specific value in the case of exact prioritization techniques class, or implicitly in the case of Relative prioritization technique class. Also, each class is divided into two subclasses based on the way they give a rank for the requirements. After that, a comparison is established between these techniques.
- *Some recommendations are documented to help project managers in the process of selecting the most suitable prioritization technique based on their project characteristics including three major directions including number of requirements, time, cost, and accuracy the project must satisfy.*
- A specific property for each proposed class and subclass had been concluded based on a perfect analysis for comparison results which supports the correctness of the novel proposed classification.

The rest of this survey is organized as follows: section 2 presents the Literature review, section 3 describes some requirements prioritization techniques distributed in the proposed classes. The comparison of requirements prioritization methods and proposed classes is presented in section 4. Finally, Section 5 provides a conclusion and future.

2. **Literature review**

In literature, there are many types of research about requirement prioritization subjects, divided into three research directions:

1- First direction: providing a systematic review of specific requirement prioritization techniques and comparing those techniques using specific criteria. One of the recent reviews was done by Hudaib et al. [7]. They presented nine requirement prioritization techniques and compared them based on specific criteria. Also, they provided a brief discussion about using data mining and machine learning in requirement prioritization. Another review is done by Bokhari et al. [8] where a systematic literature review of ten requirement prioritization techniques (AHP, Hierarchy AHP, Bubble sort, Binary search tree, Minimal spanning tree, Numerical assignment, planning game, priority groups, value-oriented prioritization (VOP) & cumulative voting|) had been provided. These techniques were compared based on time, usability, and accuracy criteria. Siddiqui et al. [9] compared two different techniques Analytical Hierarchy Process (AHP) and Planning Game (PG) based on various factors. In [54], the prioritization strategies that are now in use have several limitations, as observed by the authors. These include concerns related to requirements reliance, lack of scalability, and how to handle rank modifications during requirements evolution. Furthermore, there hasn't been any reporting on how well-suited the current methods are in challenging real-world settings. In order to find primary research that pertain to requirements prioritization and are categorized under journal articles, conference papers, seminars, symposiums, book chapters, and IEEE bulletins, they therefore suggested search terms containing pertinent keywords. 73 Primary studies were chosen from the search processes, according to the results. Thirteen journal publications, thirty-five conference papers, and eight workshop papers were produced from these investigations. Additionally, there were two contributions from each of the IEEE bulletins and symposiums. In [55], The authors' goal was to classify and identify the prioritizing criteria that have been covered in the extensive literature on software development prioritization. They presented a consolidated prioritizing criterion model after describing a thorough literature review. This work presented a classification schema that enabled researchers and practitioners to quickly find prioritizing criteria and relevant literature, in addition to providing a thorough summary of the criteria that have been explored in the literature. There are many researches in this direction in literature [10, 11, 12, 13, and 14].

2- Second direction: proposing a novel requirement prioritization technique and comparing its result with some common existing techniques. One of the research in this direction was done by Bubo & Voola [15]. They proposed a novel technique called Extensive Numerical Assignment (ENA) which acknowledges the uncertain and incomplete nature of human judgment about requirements priorities. Also, Elsherbeiny et al. [16] provided a method for predicting the priority for one requirement from another that is positively correlated to it. They conducted their experiments on a dataset of 76 stakeholders, 10 project objectives, 48 requirements, and 104 specific requirements. In the proposed way for prioritizing requirements the frequency for each requirement is computed and compared with specific threshold values to get rid of insignificant requirements and then the remaining requirements based on the mean value of the rate for each requirement. After that, they got all requirements with high ratings and computed the correlation coefficient between these requirements to get the association between the requirements so that the rating of a requirement can be predicted if a stakeholder doesn't rate based on his other ratings. Khan et al. [17] proposed a new requirement prioritization technique called the Analytic Network Process (ANP) for an interdependent requirement. They used MATLAB for simulation.

3-Third direction: Classifying some requirement prioritization techniques into specific classes as in [18] vestola classified requirements prioritization approaches into four different abstraction levels: prioritization activities, techniques, methods & processes and mentioned some studies about these levels of abstractions that presented in the previous work in requirement prioritization research field. Nine basic requirements prioritization techniques (Numeral assignment technique, Analytic hierarchy process (AHP), Hierarchy AHP, Minimal spanning tree, Cumulative voting (CV), Hierarchical cumulative voting (HCV), Priority groups, Binary priority list (BPL), and Bubble sort technique) had been presented in this work and classified into three classes: nominal scale, ordinal scale, and ratio scale. Narendhar et al. [19] presented nine techniques and classified them into three classes based on technique results. Also, they discussed some of the requirement prioritization aspects such as importance, time, cost, penalty, and risk. Hudaib et al. [20] provided an overview of the requirement process and requirement prioritization concept and compared some of the most common prioritization techniques. Also, some approaches that are used to prioritize non-functional requirements are discussed.

Based to our best knowledge, there not exist works in the literature for leading the project managers in the process of selecting the most suitable technique to prioritize requirements based on their project characteristics. In this manuscript, we classified the existing techniques into two major classes: relative and exact prioritization techniques based on the way the value of ranking is given to the requirement. *We also provide an overview of fifteen different requirements prioritization techniques that are classified under our proposed classification. Moreover, we make a comparison between methods that are related to the same subclass to analyze their strengths and weaknesses. Based on the comparison results, the properties for each proposed subclass of techniques are identified. Depending on these properties, we present some recommendations to help project managers in the process of selecting the most suitable technique to prioritize requirements based on their project characteristics.*

3. Requirement prioritization techniques

Our work focuses on the classification of the existing techniques into two major classes: relative and exact prioritization techniques. Further classifications for each class are presented as shown in Figure 1. We depend in our classification on the way the value of ranking is given to the requirement, either explicitly as a specific value in the case of the exact prioritization techniques class, or implicitly in the case of the Relative prioritization technique class.

### 3.1. Class1- Relative prioritization techniques

In this class, the rank for each requirement is given implicitly either by assigning each requirement to one specific category from different categories or by representing its relative position concerning the other requirements in the set. This class involves two types of techniques:

- **Grouping-based techniques**, in which each requirement is assigned to one specific class/group among several classes/groups, and all requirements in the same class have the same priority. Therefore, no specific value of ranking is given to the requirement.

- **Search-based techniques**, where the rank of each requirement is represented depending on its relative position concerning the other requirements, and that is why we called it relative ranking.

## 3.2. Class2 - Exact prioritization techniques

In this class, the value of ranking is given to the requirement, explicitly as a specific value. Each requirement has a specific rank. This class involves two types of techniques:

- **Absolute evaluation techniques**, where the rank is represented by giving an absolute measure of the evaluation based on the stakeholder opinion.
- **Pairwise comparison techniques**, where a preference value is given for a requirement based on pairwise comparison.

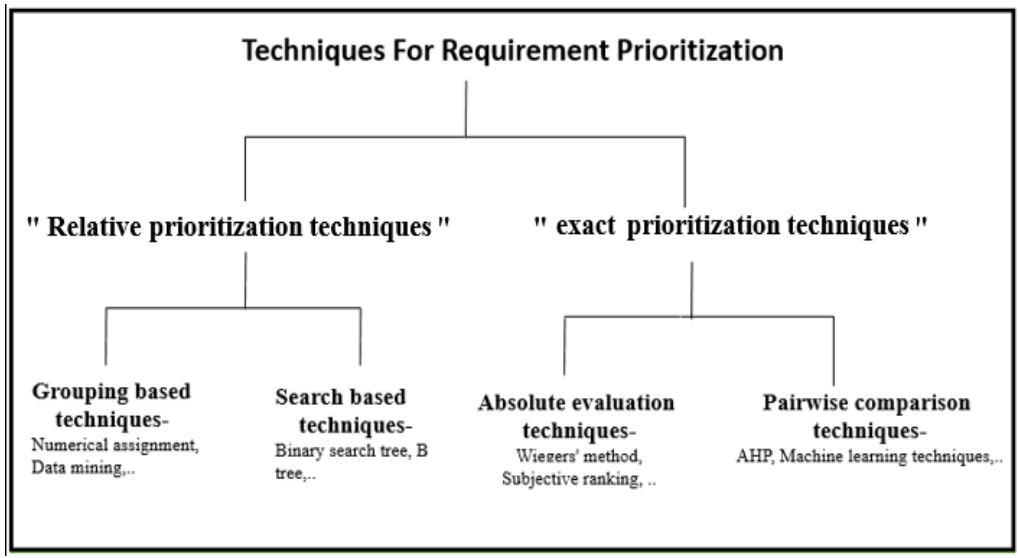

**Figure 1: Techniques for requirement prioritizations**

In this manuscript, we select the papers that are used for evaluation based on the following assessment questions:

- Does the paper proposed and/or evaluated requirements prioritization methods investigated empirically in real-life settings?
- Does the paper discussed the evaluation criteria that we used in this manuscript like ease of use, time, cost, accuracy and scale?

**Class1- " Relative prioritization" techniques**

In this class, the rank is given implicitly for a requirement either by assigning each requirement to one specific category from different categories as in grouping based techniques, and data mining techniques, or by representing its relative position concerning the other requirements in the set, as in the Search-Based techniques. One shortcoming of this class is the fact that no unique priority is assigned per requirement.

## 4.1. Grouping Based Techniques

- **Numeral Assignment Technique (NA)**

Numerical assignment [21] is one of the most common prioritization techniques, where requirements are grouped into different groups [22]. The number of priority groups varies but the most common number of groups is three [23, 24] called "critical, standard, and optional".

The specified groups must be defined clearly to allow all stakeholders to have the same understanding of each group during the prioritization process. The percentage of requirements that can be placed in each group should be restricted to prevent the stakeholders from putting all requirements in one category [25], requirements in the same group have the same priority.

Bubo & Voola [15] Have been analyzing and extending this technique to Extensive Numerical Assignment (ENA) which acknowledges the uncertain and incomplete nature of human judgment about requirements priorities, which are in turn uncertain guesses about the upcoming product and compared it with two common requirements prioritization techniques: Numerical Assignment and AHP experimentally by prioritizing the requirements of a university website system with the university students as a stockholder in the experiment. The comparison is made based on various objective and subjective measures like the number of decisions, time consumption, ease of use, attractiveness, and scalability. The results showed that ENA outperformed NA and AHP techniques.

- **MoSCoW**

The MOSCOW technique was introduced by Dai Clegg of Oracle UK in 1994 and mentioned in many types of research 19, 25, 26, and 27] it is a kind of numerical assignment technique mentioned in DSDM. It is a dynamic systems development method that works by dividing requirements into four groups called: "Must have, Should have, Could have, and won't have". Where:

-"**Must have**": means that the requirements in this group must be contained in the project (the success of the development project depends on these requirements).
- "**Should have**": means that the project would be better if it contains requirements specified in this group.
- "**Could have**": means the project will look extremely better if it contains these requirements.
- "**Won't have**": means that the requirements in this group are good but not needed in the current time, may be needed in the future. [28]

- **Planning Game (PG)**

The planning game is a feature of extreme programming [29] used with customers to prioritize features based on stories. This is a variation of the Numeral Assignment Technique, where the customer distributes the requirements into three groups, "those without which the system will not function," "those that are less essential but provide significant business value," and "those that would be nice to have."[30]

## Data mining techniques

- **Clustering based techniques**

Duan et al. [31] proposed a method that combines data mining and machine learning techniques for providing requirements prioritization using an automated clustering algorithm according to stakeholders' interests, business goals, and cross-cutting concerns such as security or performance requirements, where each requirement at the beginning put in the separate cluster then proximity score is computed between each pair of requirements by computing the probability that two requirements represent the same concept, based on the Similar terms.

This process involves specific steps before computing the similarity between two requirements like a preprocessing step where stop words are removed, and the remaining words are stemmed to their roots. This process is repeated until a stopping condition is reached which is identified as an average size of the cluster that equals seven in the case study based on the requirements of the Ice Breaker System (IBS) that is used in this paper.
In the IBS case study, the clustering algorithm generated 41 distinct clusters. After that, the method progresses by prioritizing the defined clusters using any prioritization method (here BST is used). Finally, requirements at each cluster have been prioritized based on the cluster's prioritization.

Achimugu et al. [32] are concerned with situations where the number of requirements to prioritize is large. They used the KMean clustering algorithm for solving requirements prioritization problems by using the weights of attributes of requirement sets from relevant project stakeholders as input parameters to the algorithm. In this paper, they validated their work using the RALIC dataset. The results showed the effectiveness of using this algorithm for prioritizing requirements.

RP-GWO is another algorithm that emulates the hunting behavior of grey wolves in nature to prioritize the requirements [33]. This algorithm starts with initializing the gray wolf population and then computes the number of clusters and centroids as well. After that, it computes the fitness function for each search agent. The fitness function is the distance between each wolf and each centroid. Therefore, each wolf joins the closest cluster and is a member of this cluster. The findings display that the RP-GWO performs better than the AHP mechanism by approximately (30%) in terms of average running time and dataset size.

· **Classification based techniques**

Cleland-Huang et al. [34] used the NFR classifier for identifying the cross-cutting and other architecturally significant requirements. Ice Breaker System (IBS) was used in this paper as a case study. IBS included a significant number of security, usability, look-and-feel, reliability, and extensibility requirements but the NFR classifier hadn't previously been trained for detecting reliability and extensibility requirements so, the IBS data set retrained the NFR classifier for classifying the new types of requirements. Five NFR types were prioritized on a scale of 0–5 with 5 being most important (reliability), usability and look & feel (2), and security and extensibility (1). Finally, requirements from each NFR type are prioritized based on the type of prioritization.

## 4.2. Search-based techniques

The rank for the requirement in these techniques can be expressed as its relative position concerning other requirements in the set, as in the Binary Search Tree, B tree, Bubble sort, and Minimal Spanning Tree.

· **Binary Search Tree (BST)**

A binary Search Tree [30] is an algorithm sometimes called an ordered or sorted binary tree, that is used for search as it allows fast lookup, addition, and removal of items, and can be used to implement either dynamic sets of items, or lookup tables that allow finding an item by its key and can easily be scaled to be used in prioritizing many requirements [35]. The basic approach for requirements prioritization using BST is as follows:
1- Take one requirement and put it as the root node.
2- Insert the next requirement in the appropriate position in the tree based on its priority.
3- Repeat step 2 for all the remaining requirements.

NF or better presentation, traverse through the entire BST in order and put the requirements in a list, with the least important requirement at the end of the list and the most important requirement at the start of the list.

· **B- Tree (BT)**

1. Rizwan Beg Proposed B-tree [36], which is a requirements prioritization technique that uses a balanced search tree to maintain easy searching of the requirements on one hand and to minimize the number of pairwise comparisons on the other hand which keeps the process of prioritizing a large number of requirements easy and simple. However, in this method, the requirements are assumed to be independent which affects the ability to use this technique in a situation where there are dependent requirements.

· **Bubble sort (BS)**

Bubble sort [37] was introduced in 1998 by Karlsson. It is an algorithm for ranking requirements by comparing two requirements at a time and swapping them if the two requirements are not in the correct position. The process continues until no more swaps are needed (all requirements map to their appropriate place based on their priority value) [38].

1. **Class2 - "exact prioritization" techniques**

The rank of requirement is explicitly assigned to each requirement according to a specific criterion, it can be represented by giving an absolute measure of the evaluation or by assigning a preference value to pairs of candidate requirements based on pairwise comparison.

## 5.1. Absolute Evaluation Techniques

·   **Wiegers' Method**

This method was invented by Karl Wiegers [40]. It relates directly to the value of each requirement to a customer. The priority is calculated by dividing the value of a requirement by the sum of the costs and technical risks associated with its implementation.

The value of a requirement depends on both the value provided by the client to the customer and the penalty that occurs if the requirement is missing. This means that developers should evaluate the cost of the requirement and its implementation risks, as well as the penalty incurred if the requirement is missing. Attributes are evaluated on a scale of 1 to 9. [41]

·   **Subjective ranking (SR)**

In this technique, each stockholder selects a priority value from a scale where each one provides his special opinion based on the importance of each requirement to him. This technique is done through meetings or electronic mail. The rank of each requirement was calculated after that by taking the average value of all stakeholder's priority values that they assigned to that requirement. This technique leads to conflicting priorities because all opinions have identical weights. [42]

·   **Five Whys (FW)**

In this method for each requirement, each stockholder was asked at least five "why" questions to discover whether the requirement was truly needed after just a few "whys" and then provided a rank based on these stakeholder's answers. [42]

It often happens that stakeholders want to implement a certain feature for reasons that are not founded on logical arguments or the business interests of the company, so this technique allows the analyst to determine whether the requirement is really necessary or can be canceled once the priority is determined. [43]

·   **Limited Votes (LV)**

Limited Votes is a scheduling strategy that forces reluctant stakeholders to make decisions. Each stakeholder gets a limited number of votes that can be assigned to any of the identified requirements. Multiple votes per requirement are allowed (multi-voting). The key is to provide each stakeholder with fewer votes than there are requirements. This forces the stakeholders to make decisions. If some requirement is truly crucial to them, then they can give it more than one vote; of course, that will take a vote away from some other requirement that they'd perhaps like included. [42]

· **Limited weighted votes (LWV)**

It is like the limited votes technique except that not all stockholders have the same weight (like providing higher weights for important stockholders). In this way, important stakeholders can be more influential in requirement prioritization decisions as they have higher weights than other stakeholders. This technique is concerned with identifying stockholders and their importance to the project. [42]

## 5.2. Pairwise comparison techniques

· **Analytic Hierarchy Process (AHP)**

It is one of the most commonly studied requirements prioritization, it was developed by Saaty [44] and applied to the software engineering field by Karlsson [45].
In this method, we estimate the relative importance between all unique pairs by performing a comparison to decide which requirement is more important using a scale from 1-9.

If we have n number of requirements we need to create an n x n matrix and insert the n requirements in the rows and columns of the matrix. Then for each pair of requirements, we insert their relative intensity of importance in the position where the row of the first requirement meets the column of the second one.

Meanwhile, the reciprocal values are inserted into the transposed positions. The diagonal of this matrix will be 1 since we compare the requirement with itself. Finally, we calculate the values of the resulting comparison matrix, and the relative priorities of the requirements are achieved. AHP requires $n \times (n-1)/2$ comparisons. It is worth mentioning that AHP also includes a consistency check to check the accuracy of the comparisons.

- **100-Point Method or cumulative voting (CV)**

The 100-Point Method [46] also called cost-value or cumulative voting (CV) where each stakeholder has to distribute 100 points among requirements based on their importance according to his point of view [45]. For example, if there are four requirements that the stakeholder views as an equal priority, he or she can put 25 points on each.

A hierarchical version of CV which is called HCV has been developed for dealing with the case of having multiple levels of requirements. The main problem with a CV is that it is difficult to distribute the points for a large number of requirements. When solving this problem, the amount of points is increased to avoid the scalability problem. [47]

- **Minimal Spanning tree (MST)**

It is a prioritization method which is introduced by Karlsson [39]. Minimal spanning is a directed graph that is connected minimally. This method rests on the idea that redundancy will not happen if the decisions are perfectly consistent. Therefore, if we have n number of requirements then the number of comparisons will be reduced to n-1.

The minimal spanning tree is suitable for a project that contains a large number of requirements as it is considered a fast technique due to minimizing the number of pairwise comparisons. However, it is sensitive to judgmental errors as all redundancy has been removed.

**Machine learning techniques**

Paolo Avesani et al. [48] proposed a novel framework that exploits machine learning techniques to reduce the elicitation effort in the prioritization task. This can be achieved by estimating the unknown preferences based on the other acquired ranks according to predefined prioritization criteria. The results of this technique outperformed the Analytic Hierarchy Process (AHP) method as the total amount of information that has to be acquired from the stakeholder in (AHP) increases quadratically with the number of requirements, which affects the scalability issue. This limitation has been solved with the collaboration of machine learning techniques by approximating part of the pairwise preferences for prioritizing requirements.

A set of experiments have been conducted with a group of students of the computer science faculty, using a case study extracted from a real application. The experimental results were really promising since they obtained an accurate requirements ranking with a limited elicitation effort.

The previous work has been continued [49] where a case-based framework for requirements prioritization has been adopted, called Case-Based Ranking (CBRanking). The framework was described in detail and empirical evaluations have been conducted to show the effectiveness of exploiting machine learning techniques to overcome the scalability problem.

This work described the basic steps of the prioritization process, which takes several requirements Req = {r1, r2. . . rn} as input, and computes the approximation of the target ranking as output. These steps can be summarized in three steps as follows:
- Pair sampling: This is an automatic step that selects a pair of requirements, (ri, rj).
- Preference elicitation: This step is in charge of the stakeholder. It takes a collection of pairs of requirements as input and it produces their ranks as output.
- Ranking learning: It takes in the stakeholder preference as input, and computes an approximation of the ranking function H(r). Moreover, it exploits the boosting approach and tries to estimate to rank the unknown pairs.

This method reduces the acquisition effort by combining human preference elicitation and automatic preference approximation. The results proved that the proposed framework is effective in dealing with large sets of requirements.

**Methods Exploiting Genetic Algorithm (GA)**

Gunther Ruhe et al. [50] described a method called EVOLVE+ which is based on a genetic algorithm to prioritize requirements, this method is extended from [51]. EVOLVE+ provides a quantitative analysis to support decision-making for software release. It has a significant role in minimizing the decision-making effort and increasing the accuracy of the requirements ordering as well.

Paolo Tonella et al. [52] used an Interactive Genetic Algorithm to order requirements taking into account the relative preferences elicited from the stakeholders. It aims at minimizing the amount of knowledge that has to be elicited from users which enhances the scalability of this technique over others. This algorithm has been applied to a real case study consisting of a large number of requirements, the results outperformed GA.

1. Comparison between requirements prioritization techniques

Based on the discussion given above for the 15 techniques, we compare these techniques in terms of Accuracy, Scalability, ease of use (usability), Scale, and number of requirements as shown in Table 1. This comparison done by analyzing some project manager experiments with a real Projects, and the discussed related works related to requirement prioritization that are displayed in this manuscript.

As can be seen in Table 1, grouping-based techniques such as Numerical Assignment and Moscow, have similar characteristics. All these techniques are based on prioritizing the requirements into groups, so these techniques are simple, fast, and characterized by low complexity as they have minimum computations compared to the other techniques. This subclass is suitable for projects with a large number of requirements and it has high scalability.

**Table 1. Comparison of Grouping Based Techniques**

| Method | Scale | Req.No | Complexity | Accuracy | Scalability | Usability |
|---|---|---|---|---|---|---|
| NA | Ordinal | Large | Low | Low | High | High |
| MOSCOW | Ordinal | Large | Low | Low | High | High |
| PG | Ordinal | Large | Low | Medium | High | Medium |

In the case of Search-based techniques like BST, BS, and B tree; these techniques can only show that one requirement is more important than another requirement, but not to what extent. Therefore; we can specify this subclass properties as easy to use and their speed is almost medium. Techniques in this subclass are suitable with a small or medium number of requirements, as shown in Table 2.

**Table 2. Comparison of Search Based Techniques**

| Method | Scale | Req.No | Complexity | Accuracy | Scalability | Usability |
|---|---|---|---|---|---|---|
| BT | Ordinal | Medium | Medium | High | Medium | High |
| BST | Ordinal | Medium | Medium | High | Medium | High |
| BS | Ordinal | Small | Medium | Medium | Low | High |

Regarding the absolute subclass techniques, it can be seen in Table 3 that they are usually used with a small number of requirements and the scalability rate is low since the priority is given to each requirement as an absolute value, based on the stakeholder opinion. The accuracy is medium and they are not easy to use.

**Table 3. Comparison of Absolute evaluation techniques**

| Method | Scale | Req.No | Complexity | Accuracy | Scalability | Usability |
|---|---|---|---|---|---|---|
| Wieger's | Ratio | Small | High | Medium | Low | Low |
| SR | Ratio | Small | High | Medium | Low | Low |
| FW | Ratio | Small | High | High | Low | Low |
| LWV | Ratio | Small | High | High | Low | Low |
| LV | Ratio | Small | High | Medium | Low | Low |

Methods that belong to the Pairwise techniques subclass are suitable for large requirements with high accuracy except for AHP which has high complexity and can handle a small number of requirements as it has relatively high computational complexity. However, MST is faster since it eliminates some comparisons.

Machine learning methods such as CBRank use the same technique of pairwise comparison and add some features like estimating the unknown preferences based on the other acquired ranks. Therefore, machine learning techniques can be used with a large number of requirements and give high accuracy in the shortest time.

**Table 4. Comparison of Pairwise based techniques**

| Method | Scale | Req.No | Complexity | Accuracy | Scalability | Usability |
|---|---|---|---|---|---|---|
| CV | Ratio | Large | Low | Medium | High | High |
| CBRank | Ratio | Large | Medium | High | High | Medium |
| AHP | Ratio | Small | High | Low | Low | Medium |
| MST | Ratio | Large, Medium | Low | High | High | High |

Finally, based on analyzing the previous comparison results, we can summarize the common properties for each proposed group as shown in Table 5.

**Table 5. Comparison of requirements prioritization groups**

| Method | Scale | Req.No | Complexity | Accuracy | Scalability | Usability |
|---|---|---|---|---|---|---|
| Group based technique | Ordinal | Large | Low | Low, Medium | High | High, Medium |
| Search based technique | Ordinal | Small/Medium | Medium | High, Medium | Low, Medium | High |
| Pairwise based techniques | Ratio | Large except AHP | Low, Medium, except AHP | High, Medium except AHP | High Except AHP | High, Medium |
| Absolute Based technique | Ratio | Small | High | High, Medium | Low | Low |

From the previous discussion, we can provide some recommendations to help project managers in the process of selecting the best requirement prioritization method based on the project's nature and characteristics as illustrated in the following points:

If the project contains a large number of requirements, then any method from Group group-based techniques subclass will be the suitable choice if you need a fast and easy-to-use method but if you need better accuracy you can use some methods from pairwise-based techniques except AHP.

If the project contains a small number of requirements, then any method from any subclass will satisfy your purpose but AHP from pairwise comparison techniques subclass will be more suitable if you do not need high accuracy. For better accuracy, you can use any method from the absolute evaluation techniques subclass.

If your project contains a medium number of requirements, then BST and BT from the search-based techniques subclass will be suitable.

## 7. Conclusions and future work

In this survey, we discussed various types of requirements prioritization techniques and classified them into two general classes according to the way the rank is expressed for each requirement. After that, a comparison between these techniques was presented based on some parameters. We can say that there is no technique to be considered the best one, each one has its pros and cons. However, some techniques are more suitable than others depending on the nature and characteristics of the project. Therefore, in this survey, we provided some recommendations to guide project managers to the most suitable requirement prioritization method based on their project nature.

In the future, we intend to investigate more prioritization techniques and compare these techniques against the other techniques discussed in this paper. Also, more characteristics will be take into account for selecting the suitable requirement prioritizing technique.

## Acknowledgement

This work was supported by the Ministerio Español de Ciencia e Innovación under project number PID2020-115570GB-C22 MCIN/AEI/10.13039/501100011033 and by the Cátedra de Empresa Tecnología para las Personas (UGR-Fujitsu).## References